\documentclass[12pt, a4paper]{article}
\usepackage[font=footnotesize]{caption}
\usepackage[utf8]{inputenc}
\usepackage{graphicx}
\usepackage{authblk}
\usepackage{booktabs}
\usepackage{graphicx}
\usepackage{verbatim}
\usepackage{url}
\usepackage{caption}
\usepackage{subcaption}
\usepackage{indentfirst}
\usepackage{url}
\title{Retrieving Hierarchies}
\author{Alexandre Benatti and Luciano da F. Costa}

\date{
S\~ao Carlos Institute of Physics - DFCM \protect\\
University of S\~ao Paulo \protect\\
P.O. Box 369, S\~ao Carlos, S.P. \protect\\
13560-970 Brazil \\ \vspace{0.5cm}
\emph{10th Feb., 2022}
}

\begin{document}
\maketitle

\begin{abstract}
Several real-world and abstract structures and systems are characterized by marked hierarchy to the point of being expressed as trees. Because the study of these entities often involves sampling (or discovering) the tree nodes in a specific order that may not correspond to that originally shaping the tree, reconstruction errors can be obtained. The present work addresses this important problem based on two main resources: (i) the adoption of a simple model of trees, involving a single parameter; and (ii) the use of the coincidence similarity as the means to quantify the errors by comparing the original and reconstructed structures considering diverse sampling error probability and extent. Several interesting results are described and discussed, including the fact that the average and standard deviation values of the reconstruction errors depend only moderately on the extent of the errors as well as on the types of trees. At the same time, it is identified that the relative reconstruction accuracy substantially decreases markedly with the error probability, with larger reconstructions accuracy relative variations being observed for the smallest values of that probability.
\end{abstract}

\section{Introduction}\label{sec:Introduction}

\begin{figure}[!ht]
  \centering
    \includegraphics[width=0.65\textwidth]{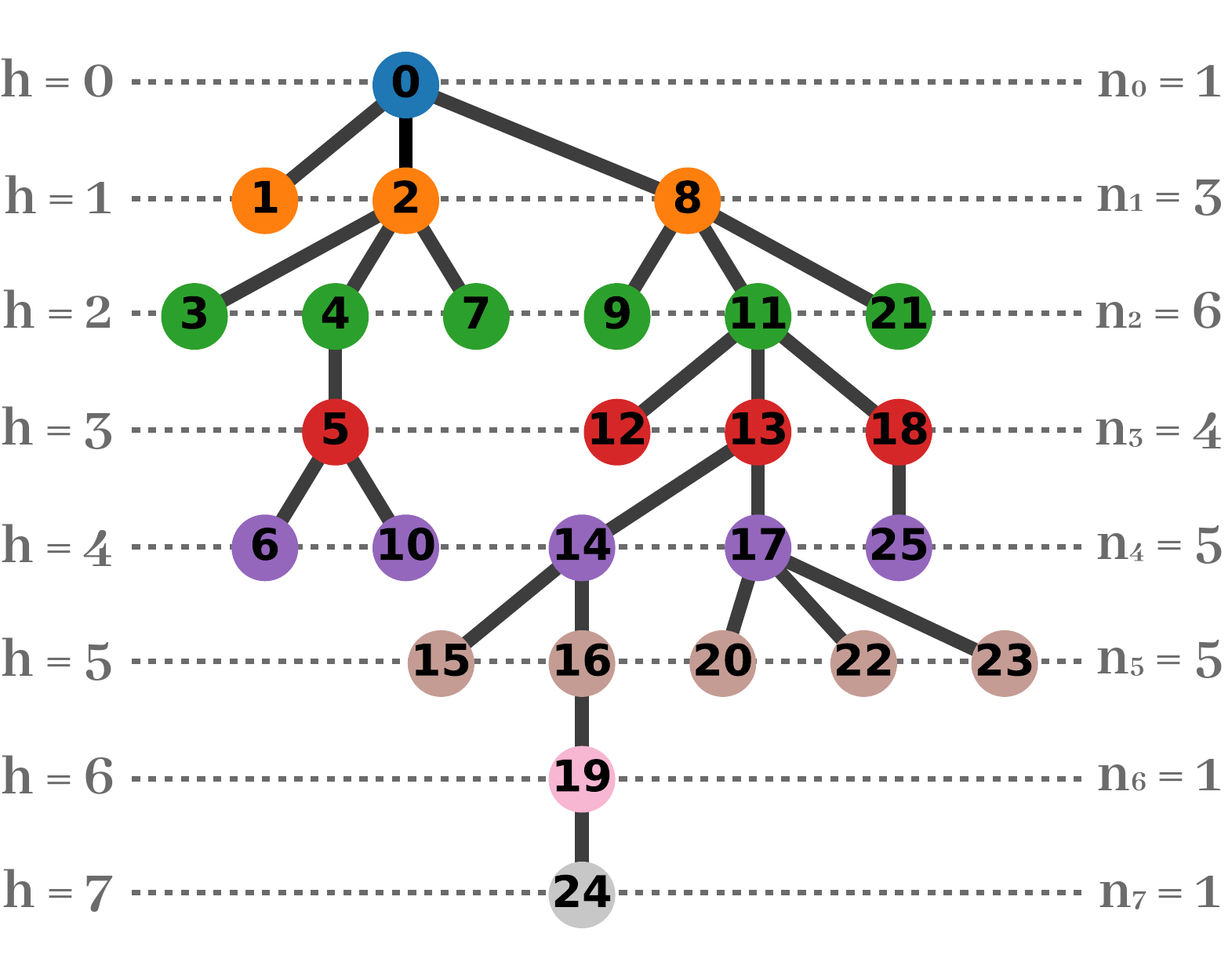}
  \caption{Example of a tree with 7 levels, also indicating the respective number of hierarchical levels ($H$) and the number of branches per level ($n_h$).}
  \label{fig:tree_1}
\end{figure}

The physical world is characterized by an impressive diversity of structures and dynamics. Among the several possible organizations to be found, \emph{hierarchies} represent a particularly interesting type of structure, being directly related to \emph{trees}, the latter corresponding to connected graphs starting at a single node (the root) and then branching successively without loops along hierarchical levels, as illustrated in Figure~\ref{fig:tree_1}.

Hierarchical organizations and trees play a particularly important role in scientific modeling because several real-world entities involve or even are completely determined by a respective hierarchical structure.  Examples of intrinsically hierarchical physical structures include but are by no means limited to roots and branches of trees, vascularization, neuronal cells, among many other possibilities.  At the same time, several abstractions underlying modeling also present hierarchical organization, including phylogenetics, taxonomies, archaeological chronology/stratigraphy, classifications, etc.  Hierarchy plats such an important  role in the modeling and understanding real-world and abstract systems that even non-hierarchical systems are often summarized in hierarchical manner, such as in terms of minimum spanning trees (e.g.~\cite{dussert1986minimal}).

It is thus hardly surprising that substantial attention has been dedicated to the study of hierarchies, including several developments aimed at characterizing, studying, modeling, and generating hierarchical structures (e.g.~\cite{yang2005similarity, emmert2005classification,pelillo1999matching, robson2021structure, onnela2004clustering}). In particular, research aimed at understanding how hierarchical structures can be generated can provide important basic subsidies for better understanding existing hierarchies.  For instance, the branched structure of a given type of plant root can be better understood provided we know how it typically arises in nature.  Additional examples of real-world related problems include ontologies, phylogenetic structures, as well as semantic structures. 

Hierarchical structures can be generated in several manners, including the situation in which new entities are sampled in a given order and progressively incorporated into a respective reconstruction, e.g.~while considering the \emph{overlap} or \emph{similarity} between the properties of the new entity and those already incorporated into the current hierarchical structure.  

\begin{figure}[!ht]
  \centering
   \includegraphics[width=0.4\textwidth]{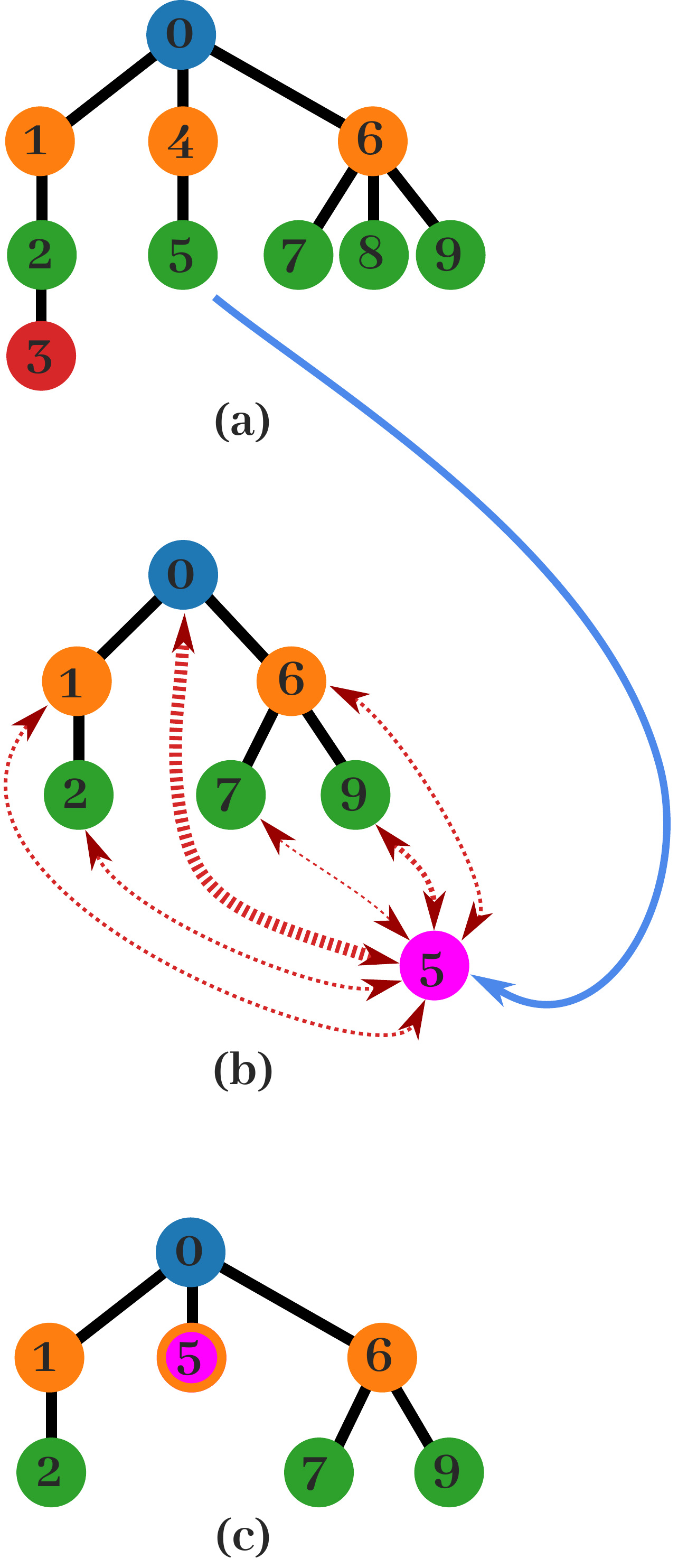}
  \caption{Illustration of reconstruction of an original reference hierarchy (a) by progressive incorporation of new sampled entities (nodes).  A newly sampled entity, shown as the magenta node in (b), has its properties compared to those of all the nodes in the current network, being linked to the node yielding the largest similarity (c). The dashed arrows indicate the similarities between the new and already existing nodes.}
  \label{fig:tree_2}
\end{figure}

Figure~\ref{fig:tree_2} illustrates the progressive reconstruction of an original reference hierarchy (a) by incorporation of successively sampled (or discovered) new entities (node in magenta).  The properties of the new node (b) are compared to those of all the nodes already available in the current tree, and the maximum pairwise similarity is identified. The new entity is then respectively linked to node 0 (c), to which it is likely most similar.  Given the great similarity between nodes 5 and 4, and nodes 4 and 0, by transitivity we have that node 5 will also be similar to node 0. This situation, in which one of the hierarchical levels along one of the tree paths (e.g.~$0,4,5 \rightarrow 0,5$) is missed because of the sampling order, constitutes the main reason for errors in the hierarchies reconstructions. 

A related question of critical interest and importance concerns to which extent different orders of sampling an original hierarchy, characterized by their various respective characteristics, can influence the respectively reconstructed trees.  In our study, it is assumed that the existing tree is never reorganized other than by the inclusion of incoming nodes.
This assumption, implying an incremental hierarchy retrieving approach, is aimed at modeling situations in which the currently available hierarchy has already been established as a reference, being unlikely to undergo major changes.

It is important to keep in mind that hierarchy retrieval depends not only on the order in which the new nodes are sampled (or discovered), but also on the original hierarchy, the features that characterize the nodes, as well as the adopted similarity metric.  In the present approach, the sampling procedure considers two related errors: (a) the probability of changing the order of the currently sampled node; and (b) the extent of this error, which involves swapping the node order with an extent (or distance) $\delta$.

The present work aims at studying this interesting and relevant problem, using the recently proposed coincidence methodology~\cite{da2021further,costa2021similarity,CostaCCompl}.  In addition, we resource to a simple but effective method for generating trees with diverse properties (e.g.~number of levels and number of nodes per level) which involves a single parameter $\gamma$ controlling the branching tendency.

Several interesting results have been reported and discussed, including the tendency of the reconstruction errors, as gauged by the average and standard deviation of the obtained coincidence values, to vary only moderately with the type of hierarchy (controled by the respective parameter of the adopted model) and with the extent of order sample error.  However, the sensitivity of the average accuracy has been verified to vary more strongly with relatively small values of the probability error, increasing less markedly thereafter.

We start by presenting a review, in a non-exhaustive manner,  some works related to hierarchical structures, and then follow by presenting the concepts of coincidence similarity, a simple but versatile model for generating hierarchies, and the problem of reconstructing hierarchies by sampling.  The results are then presented and discussed, including several interesting findings such as the relatively independence of the average reconstruction accuracy respectively to the type of trees and error extent.

\section{Related Works}
Many works have been dedicated to hierarchical models and their applications. In order to develop a solution to the two-tree matching problem, in \cite{pelillo1999matching} the authors report a formal approach to matching hierarchical structures by constructing an association graph.

In another work~\cite{mones2012hierarchy}, a measurement was proposed to convey the essential characteristics of the structure and hierarchy-related properties in a complex network. This measurement is based on the generalization of the concept of centrality, ranking nodes according to their impact on the whole network. In this same work, a visualization procedure was proposed for large complex networks, used to obtain a global qualitative image of the hierarchical nature of the network.

Several studies have suggested methods for building networks, trees, and hierarchical structures (e.g.~\cite{onnela2004clustering, bryant2001constructing, banderier2002generating}) for purposes such as studying its characteristics. Other studies have been dedicated to the classification and characterization of hierarchical structures (e.g.~\cite{yang2005similarity, emmert2005classification, banderier2002generating, stadler2018statistical}).

Similarity concepts have also been considered while studying hierarchical structures. In \cite{lakkaraju2008document} the authors propose a method to identify the similarity between documents based on a conceptual tree of these documents. In \cite{liu1999approximate}, an approach for comparing shapes is described, intended to find the best match between a pair of contours.

Works studying hierarchical network models have also been reported. As an example, \cite{robson2021structure}, identify measurements that can be used to distinguish between hierarchical and non-hierarchical networks. It was described that the lack of robustness and the hierarchical structure tend to be correlated.

\section{Basic Concepts}

Similarity measures are widely used in science and technology (e.g.\cite{mirkin, vijaymeena, akbas, costa2021similarity}), being employed to determine how much two mathematical objects are related or similar. For example, the similarity between strings can be estimated based on the characters that make up each string~\cite{vijaymeena}.  Similarity between sets of objects is also often considered as the means for data classification and clustering (e.g.~\cite{shapebook,mirkin}).

There are several alternative approaches to defining similarity, one of the most common and widely used in data analysis being the cosine distance~\cite{akbas, xia2015learning, luo2018cosine}.  This measure is defined as the cosine of the smallest angle between two vectors divided by the norms of these vectors.  The Jaccard index (e.g.~\cite{Jaccard1, Loet, da2021further}) is frequently adopted for quantification of the similarity between sets, based on the concept of set cardinality.

The \emph{Coincidence Index}~\cite{costa2021similarity,da2021further,CostaCCompl} has been described as a  measure to determine the similarity between virtually any type of mathematical entity, taking into account both the Jaccard and the Interiority (or overlap~\cite{vijaymeena}) indices.  This approach is motivated by the relative interiority between the compared sets not being captured by the Jaccard index~\cite{da2021further}, as well as by the need to generalize the Jaccard index to real-valued structures, including possibly negative values.  In the present work we consider the similarity indices respectively to real, but exclusively non-negative values.

In particular, we apply the \emph{Coincidence Index} to determine the similarity between two trees (or hierarchies) $X$ and $Y$. Thus index can be defined as corresponding to the product between the Jaccard and Interiority indices, i.e.: 

\begin{equation}
    \mathcal{C}(X,Y) = \mathcal{J}(X,Y) \ \mathcal{I}(X,Y),
\end{equation}

where $\mathcal{J}(X,Y) $ and $\mathcal{I}(X,Y) $ are the \emph{Jaccard} and \emph{Interiority} indices, respectively.  

\begin{figure}[!ht]
  \centering
   \includegraphics[width=0.5\textwidth]{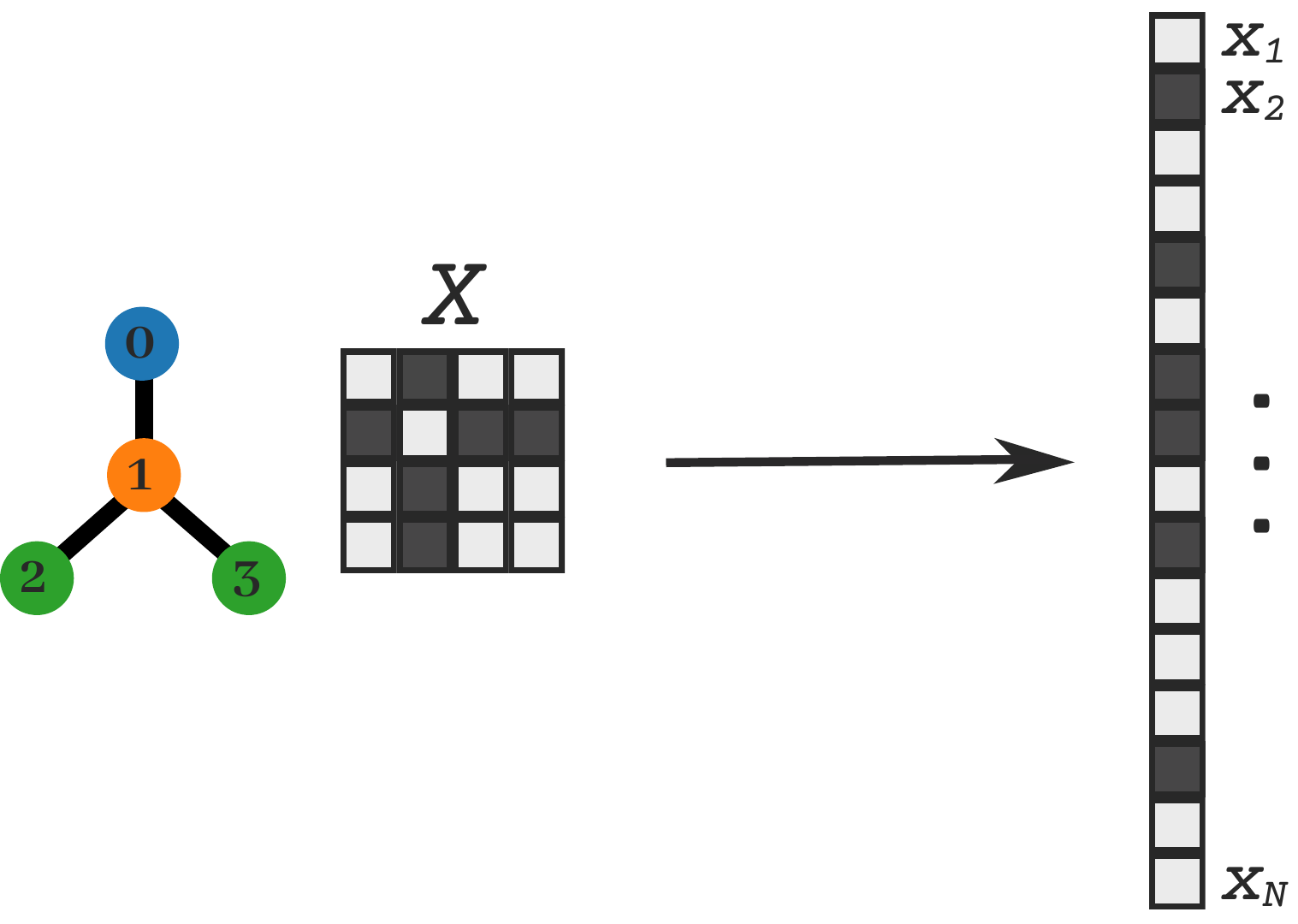}
  \caption{Example of vectorization of the adjacency matrix corresponding to a simple tree into a column vector with $N$ elements.}
  \label{fig:vectorization}
\end{figure}

The \emph{Interiority Index} ~\cite{costa2021similarity} is aimed at expressing how much one of the two sets is contained in the other set, and vice versa. The \emph{Interiority Index} between two multisets (e.g.~\cite{Hein, Knuth, Blizard, Blizard2, Thangavelu, Singh}) $X$ and $Y$ can be written~\cite{costa2021similarity} as:

\begin{equation}
    \mathcal{I}(X,Y) = \frac{\sum_i min\lbrace|x_i|,|y_i|\rbrace}{min\lbrace \sum_i|x_i|,\sum_i|y_i|\rbrace},
\end{equation} 

where $x_i$ and $y_i$ are the elements (taken as multiset multiplicities) of the trees $X$ and $Y$, understood to correspond to the \emph{vectorization} of the respective adjacency matrices representing the two trees to be compared, as illustrated in Figure~\ref{fig:vectorization}.

Observe that $0 \leq \mathcal{I}(X,Y) \leq 1$.

The \emph{Jaccard index} as a measurement of similarity between two multisets (e.g.~\cite{Jac:wiki}) $X$ and $Y$ can be expressed as:
\begin{equation}
    \mathcal{J}(X,Y) = \frac{\sum_i min\lbrace|x_i|,|y_i|\rbrace}{\sum_i max\lbrace|x_i|,|y_i|\rbrace}.
\end{equation}

with $0 \leq \mathcal{J}(X,Y) \leq 1$, so that $0 \leq \mathcal{C}(X,Y) \leq 1$. 

Observe that an adapted version of the above equation~\cite{costa2021similarity,da2021further} would be required in case the features can take negative values.

The coincidence similarity index has been found to implement a particularly strict quantification of the similarity between any two mathematical structures~\cite{costa2021similarity,CostaComparing}, being successfully applied for translating datasets into respective networks (e.g.~\cite{CostaCCompl,domingues2022city,costa2022similarity}).  Another advantage of
adopting the coincidence similarity in the present work is that it will immediately apply in cases involving hierarchies with weighted links, in addition to the binary connections characterizing the present approach.

\section{Methodology}\label{sec:Methodology}

In this section we present the simple method for generating hierarchies as well as characterize the problem of reconstructing hierarchies as developed in the current work.

\subsection{A Simple Model for Generating Hierarchies}

In order to synthesize trees having $N$ nodes with varying properties, we developed a mathematic-computational model that requires just one parameter $\gamma$, which controls how branched the trees.

\begin{figure*}[!ht]
  \centering
    \includegraphics[width=.32\textwidth]{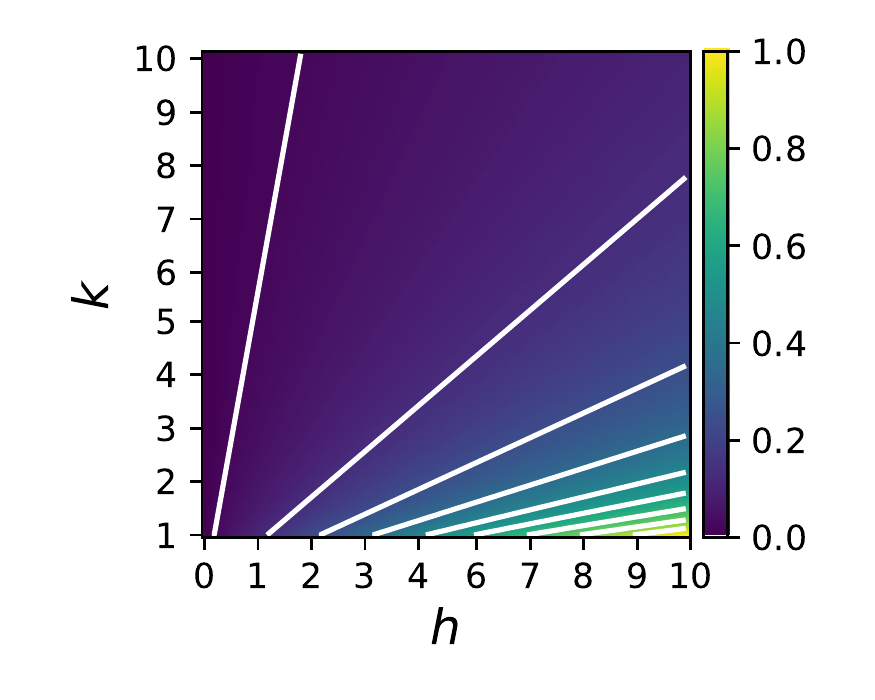}
    \includegraphics[width=.32\textwidth]{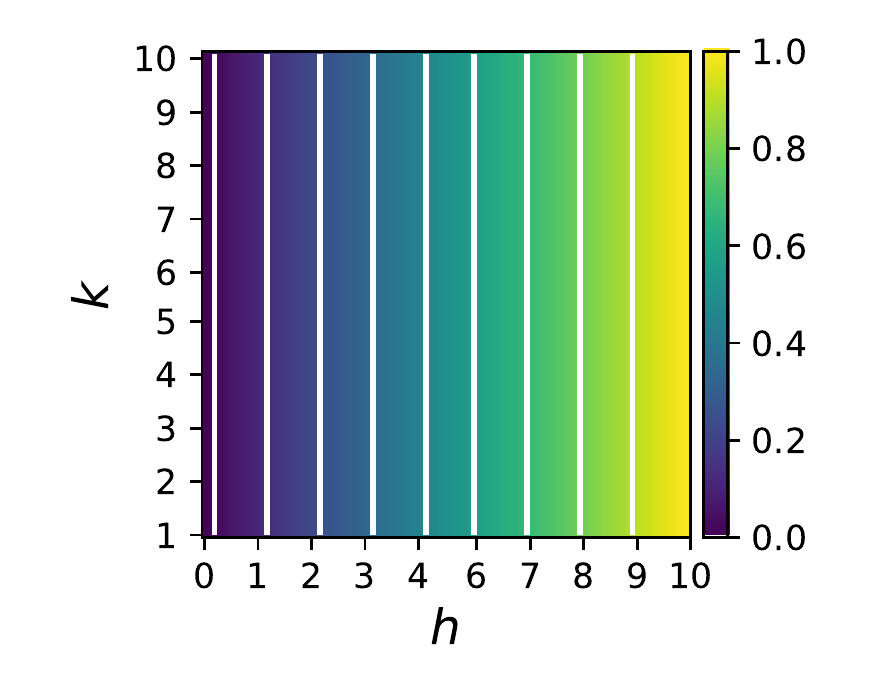}
    \includegraphics[width=.32\textwidth]{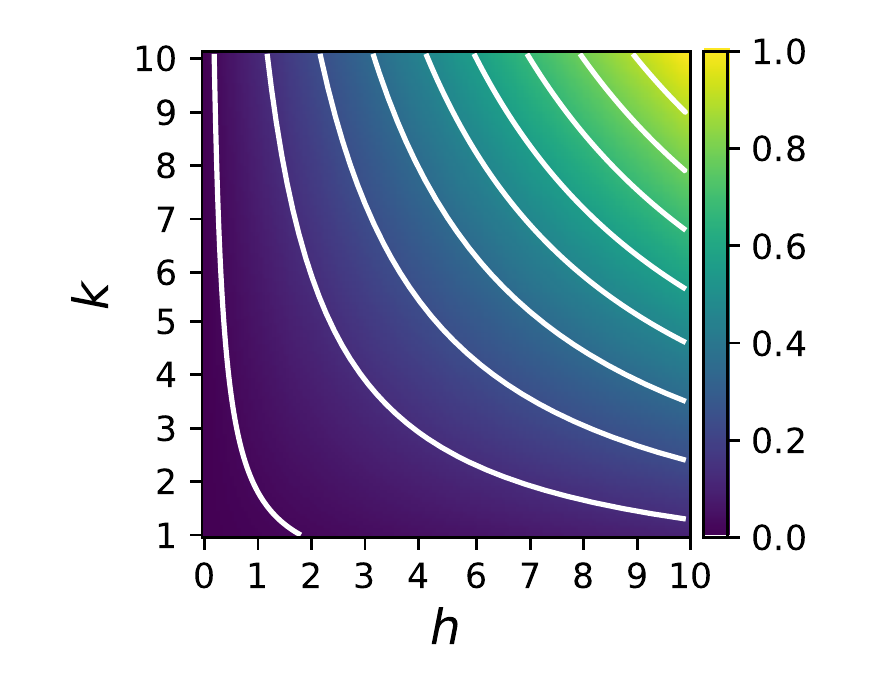} \\ 
    (a) \hspace{3.7cm} (b) \hspace{3.7cm} (c) \\
  \caption{Visualization of the probability surfaces in terms of heatmaps and level-sets in terms of $h$ and $k$ obtained for $\gamma =-1$ (a) , 0 (b) , and $-1$ (c).  For simplicity's sake, the probability values have been scale within the interval $\left[ 0, 1\right]$.}
  \label{fig:curve}
\end{figure*}

After fixing $N$ and $\gamma$, new elements are incorporated in a specific order, and each new element links only to one of the existing nodes in the hierarchical tree. The connection of the new element with some of the already existing nodes, indexed by $i$, is done randomly with the connection probability specified as:

\begin{equation}
    p_i = \frac{(h_i+1)k_i^\gamma}{\sum_j (h_j+1)k_j^\gamma},
\end{equation}

where $k_i$ is the number of links of the element $i$ and $h_i$ is its level in the  hierarchy (with the hierarchical level starting in $0$). Figure~\ref{fig:curve} illustrates the above probabilities respectively to $\gamma =-1$ (a) , 0 (b) , and $-1$ (c).

\begin{figure*}[!ht]
  \centering
    \includegraphics[width=1\textwidth]{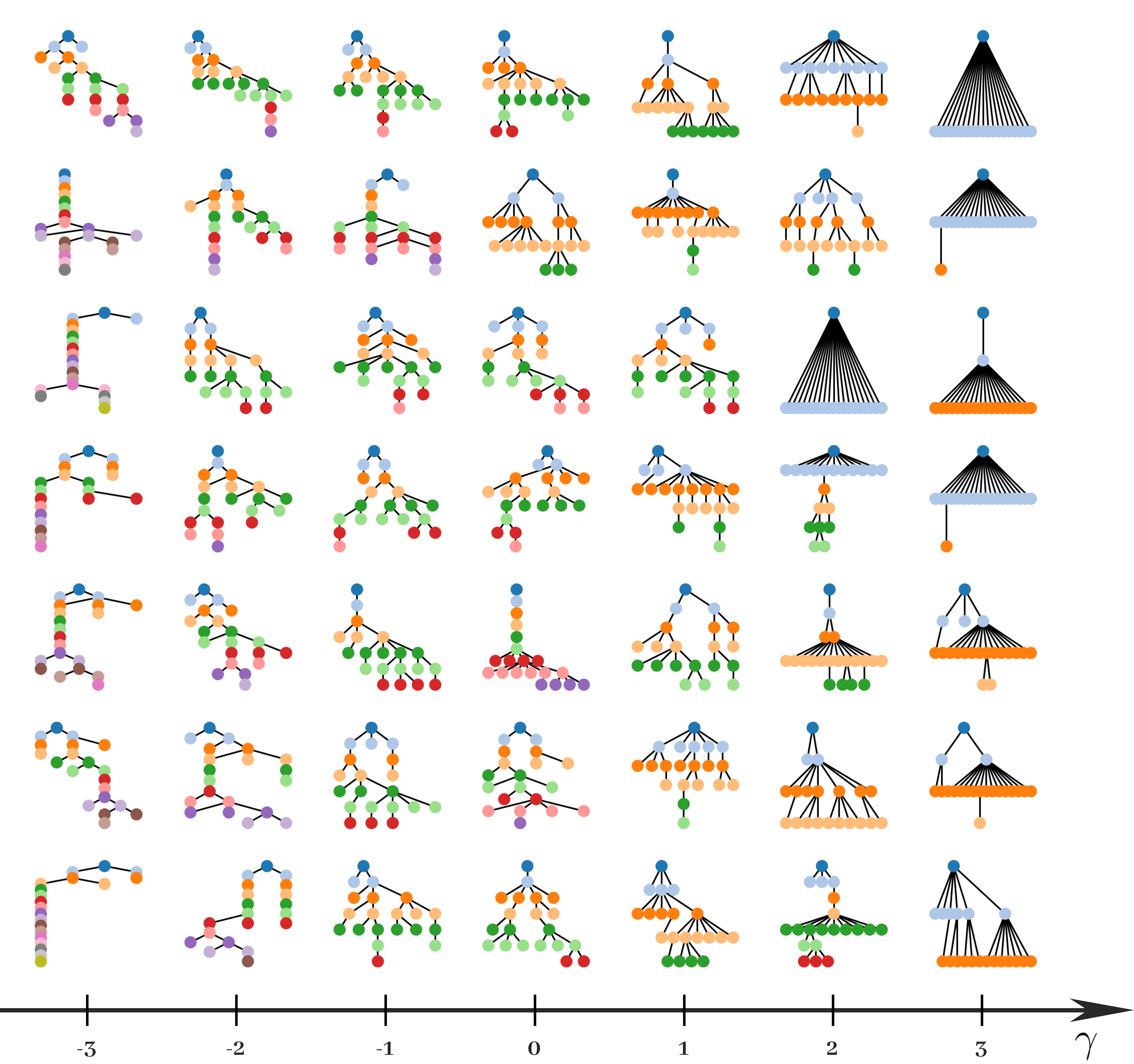}
  \caption{Examples of generated hierarchies as a function of $\gamma$, with $N=20$ nodes. More chained trees are obtained for smaller values of $\gamma$, with the number of branches increasing with that parameter.  Given that $N$ has been fixed, trees with larger number of branches will tend to have fewer hierarchical levels.}
  \label{fig:gamma}
\end{figure*}

Figure~\ref{fig:gamma} shows examples of hierarchical structures generated by the proposed model for $N=20$ and different values of $\gamma$ between $-3$ and $3$.  Observe the progression from more linear trees obtained for the smaller values of $\gamma$ to more intensely branching observed for the larger values of $\gamma$.

After the hierarchy has been obtained by using the method described above, it is necessary to associate respective features to each node, so that it becomes possible to link nodes based on the similarity between these features~\cite{CostaAmple}.  In order to do so, we start with a set $\textbf{A}$ with $m$ features: $\textbf{A} = [a_1, a_2 \dots a_m]$, where $a_i$ are integer values. Each element will receive $n$ (with $n<m$) aleatory features, with $n \beta$ of them coming from set $\textbf{A}$ and $n (1-\beta)$ coming from its \emph{parent} in the hierarchy tree. The rate $\beta$ (with $0 \leq \beta \leq 1$) determines the mix of features between the elements in this model.

Observe that, as a consequence of the method adopted for incorporating features (content) into each node, these nodes will result similar to the parent node from which they derive (one hierarchy higher), but still remaining partially distinct among themselves.  In other words, each node has a portion of its features shared with the node from which its derived, while the remainder portion is specific to itself.  This property is coherent with the concept of hierarchies, in which child nodes inherit properties, but also have specific distinguishing features (e.g.~\cite{CostaAmple}), making each of them a specific non-exchangeable entity.

\begin{figure*}[!ht]
  \centering
    \includegraphics[width=0.9\textwidth]{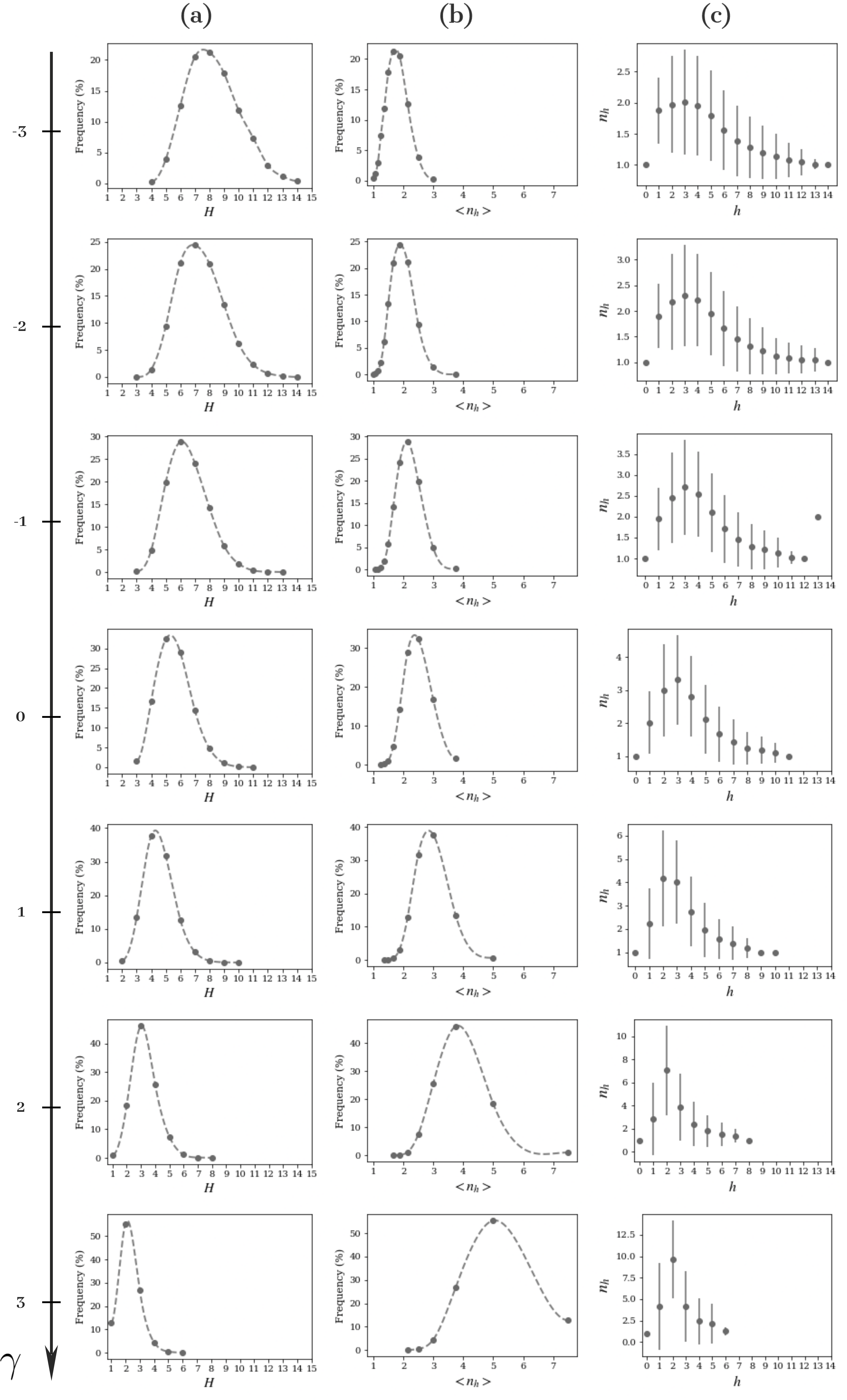}
  \caption{Properties of the trees generated by the proposed methodology in terms of $\gamma$, for $N=15$: (a) relative frequency histogram of the number of resulting hierarchical levels $H$; (b) relative frequency of the average number of nodes per level $\left< n_h \right>$; and (c) average $\pm$ standard deviation of the number of nodes per level.  These results were obtained from 10,000 realizations for each value of $\gamma$.}
  \label{fig:characterization}
\end{figure*}

Figure~\ref{fig:characterization} depicts three important topological properties of trees generated by the proposed method respectively to several values of $\gamma$.  The total number of nodes is henceforth kept fixed as $N= 15$.

Regarding the number of hierarchical levels $H$ -- Figure~\ref{fig:characterization} (a), which is among the most important property of a tree, a gaussian-like distribution can be observed respectively to each of the considered parameters configurations. The most frequent value of $H$ (abscissa of the density peaks) decreases steadily with $\gamma$. This is a direct consequence of the fact that as more branches per level are favored by larger values of $\gamma$, the number of levels tends to decrease so as to keep $N$ constant (see Figure~\ref{fig:gamma}).  At the same time, and for similar reasons, the width of the obtained densities also tends to decrease.

The average number of nodes per hierarchical level $\left< n_h \right>$, shown in Figure~\ref{fig:characterization} (b), also presents a gaussian-like profile respectively to each value of $\gamma$.  Contrariwise to the number of hierarchical levels, the most frequent values of $\left< n_h \right>$, as well as the respective width, tend to increase with $\gamma$.  This tendency is accounted for by the fact that more nodes are incorporated at each level for larger values of $\gamma$, while $N$ is kept fixed.

The distribution of the number of nodes per level, shown in Figure~\ref{fig:characterization} (c), resembles log normal-like profiles, with the peak abscissa positions and the distribution widths both decreasing with $\gamma$.  

In case tree configurations more specific than can be controlled by the parameter $\gamma$, i.e.~with less variance of properties are required, it is possible to incorporate a filtering stage after the generation of the trees, to select only the tree configurations with properties (e.g.~$H$ or $n_h$) falling within specified intervals.

\subsection{Reconstruction of Hierarchies} 

In this work, we study how the accuracy in the reconstruction of the hierarchical trees varies for different sampling orders. In this process, we considered the coincidence index between the adjacency matrices of the original and reconstructed trees in order to quantify the reconstruction accuracy.

Figure~\ref{fig:tree_2} illustrates the adopted procedure for reconstructing the hierarchies considering diverse sampling orders of the elements. The tree is reconstructed, one element at a time, defining its position in the hierarchy by connecting to the most similar element already existing in the network.

In order to sample the nodes from the original tree, we select a fraction $p$ of the elements to be dislocated by $\delta$ positions around its initial positions.

After these reconstructions have been obtained, we can compare it with the original hierarchy tree by using the coincidence similarity index between the adjacency matrices respective to those two graphs. 

\begin{figure}[!ht]
  \centering
   \includegraphics[width=0.8\textwidth]{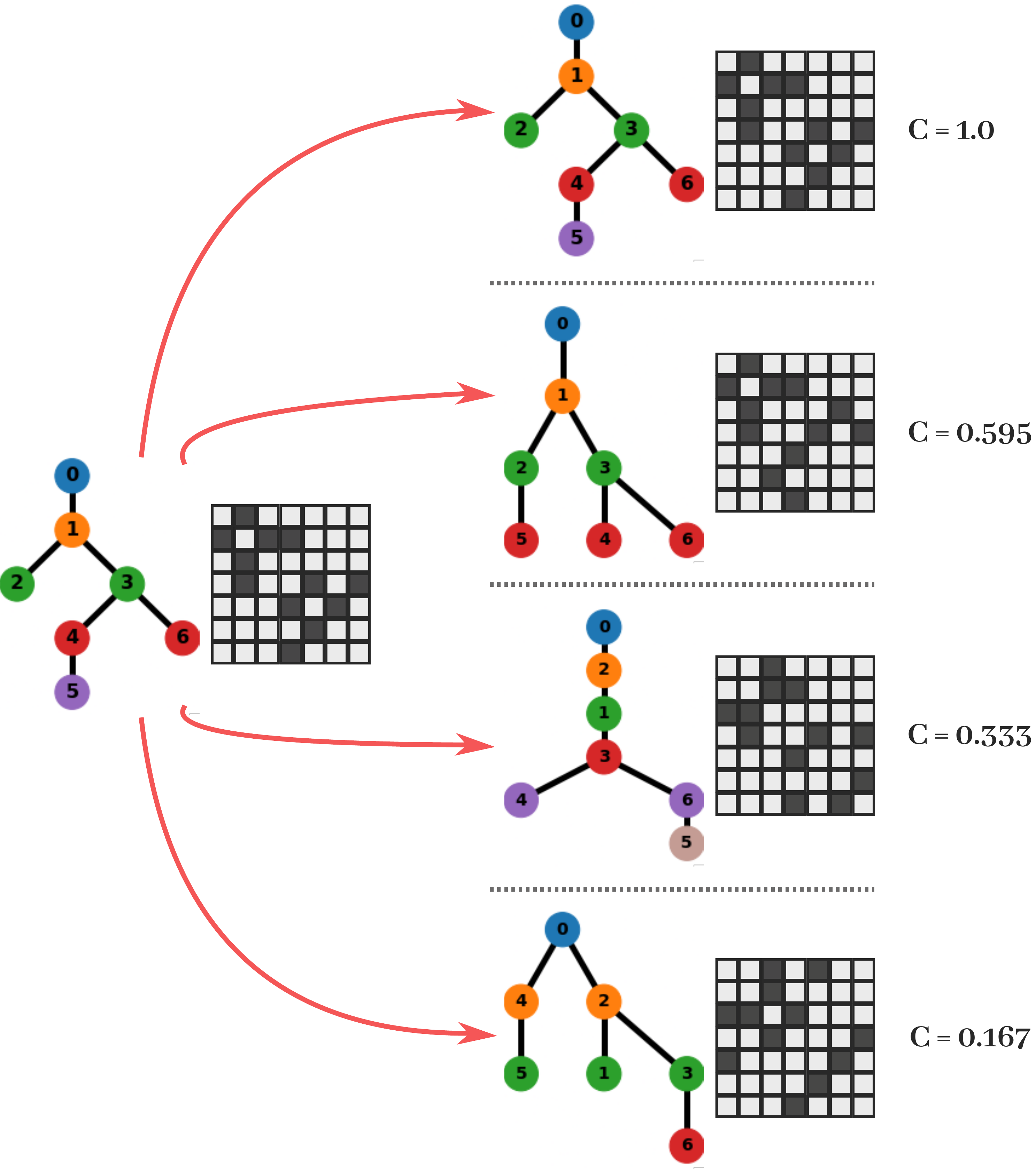}
  \caption{An original hierarchy, corresponding to the tree on the left-hand side of the figure, and four retrievals with varying accuracy. The first reconstruction results are identical to the original tree, characterized by a coincidence value $C=1.0$. Three other reconstructions with smaller accuracy are also shown, with decreasing values of $C$. The adjacency matrices corresponding to each of the trees in this figure are also shown respectively.}
  \label{fig:reconstruction}
\end{figure}

Figure~\ref{fig:reconstruction} illustrates an original hierarchy, corresponding to the tree on the left-hand size, and four respective reconstructions with decreasing accuracy.  The adjacency matrices respective to each tree are also presented respectively in the figure. The first reconstruction (top) is exact, being characterized by $C=1.0$.  Three additional reconstructions with increasing errors, therefore implying in successively smaller values of $C$.

It is interesting to observe that the finite number of nodes in the considered trees implies a discretization of the possible values of $C$ to be obtained. In this case, with 7 nodes, we have only 6 links, so that any node change implies a relatively large change inaccuracy.  The number of discrete coincidence values increases steadily with the number of nodes so that a substantially better resolution is obtained for the $N=15$ nodes adopted for the experimental results in the present work.

\section{Results and Discussion}\label{sec:Results} 

In this section, we report the results and corresponding discussions regarding the accuracy while retrieving hierarchies in the presence of sampling order errors with probability $p$ and extent $\delta$. The types of hierarchies, in the sense of the number of involved levels and nodes per level, are specified by the parameter $\gamma$ of the proposed model. More specifically, we start with a tree generated by the model for varying values of $\gamma$, which is understood as the original tree to be retrieved. Then, each of the nodes of this tree is sampled in a specific order, in presence of errors controlled by $p$ and $\delta$. Statistics of the obtained errors, as quantified by the coincidence similarity index for non-negative values are then obtained, presented, and discussed.

\begin{figure*}[!ht]
  \centering
    \includegraphics[width=1\textwidth]{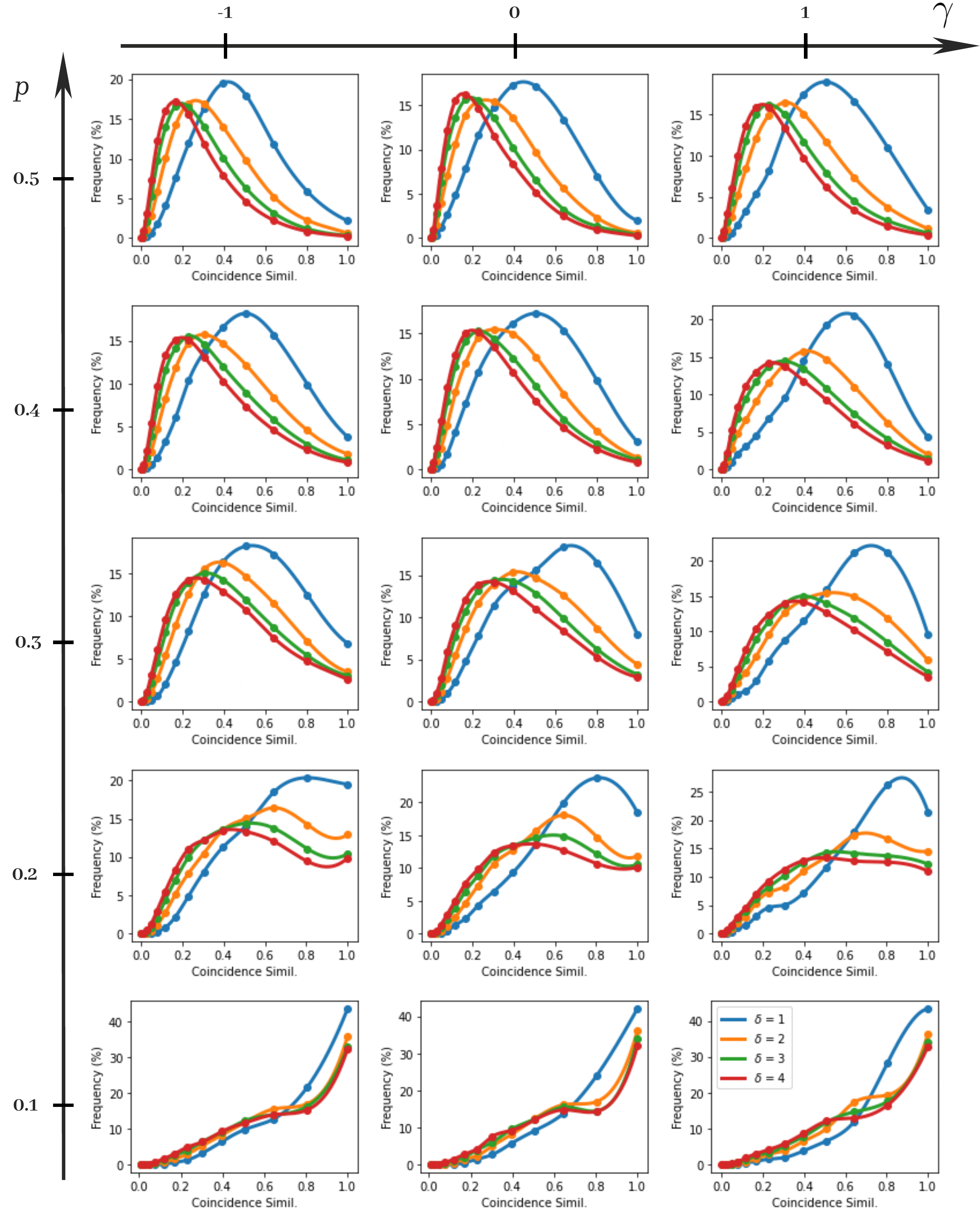}
  \caption{Relative frequency histograms of the coincidence similarity values respective to several combinations of the parameters $p$ and $\gamma$, with $N=15$ in all cases. For each of these configurations, 30 trees were generated by using the adopted model, and 4,000 different sampling orders were randomly considered,  leading to the values shown in this figure.  Five curves are shown respectively to each of the considered $\gamma$ values, corresponding to   $\delta =1, 2, 3,$ and $4$.}
  \label{fig:pXgamma}
\end{figure*}

Figure~\ref{fig:pXgamma} presents the relative frequency of the coincidence similarities obtained for several configurations of the parameters $p$ and $\gamma$. It is interesting to observe that only the similarity values corresponding to the discrete points marked along the curves were experimentally obtained, being interpolated only for the sake of enhanced visualization and comparison between the obtained profiles.  

For the smallest value of $p$, i.e.~$p=0.1$, we observe high values of similarity for every adopted $\delta$ and $\gamma$. This means that accurate reconstructions of the original hierarchies were often obtained for this probability error, with a peak near $100\%$.  However, it is important to keep in mind that relatively large reconstruction errors (i.e.~small coincidence similarity values) can be obtained, though less likely, even for this small probability error. Interestingly, the curves obtained for the different values of $\delta$ are mostly similar, except for that respective to $\delta=1$.  

For the other considered probability values $p$, the peaks of the curves obtained for $\delta = 1,2,3,4$ tend to shift from the left to right, indicating a monotonic decrease in the tree reconstruction accuracy. At the same time, for each fixed value of $\delta$, the coincidence curve also tends to shift from right to left as $p$ is increased, or $\gamma$ is decreased.

\begin{figure*}[!ht]
  \centering
    \includegraphics[width=1\textwidth]{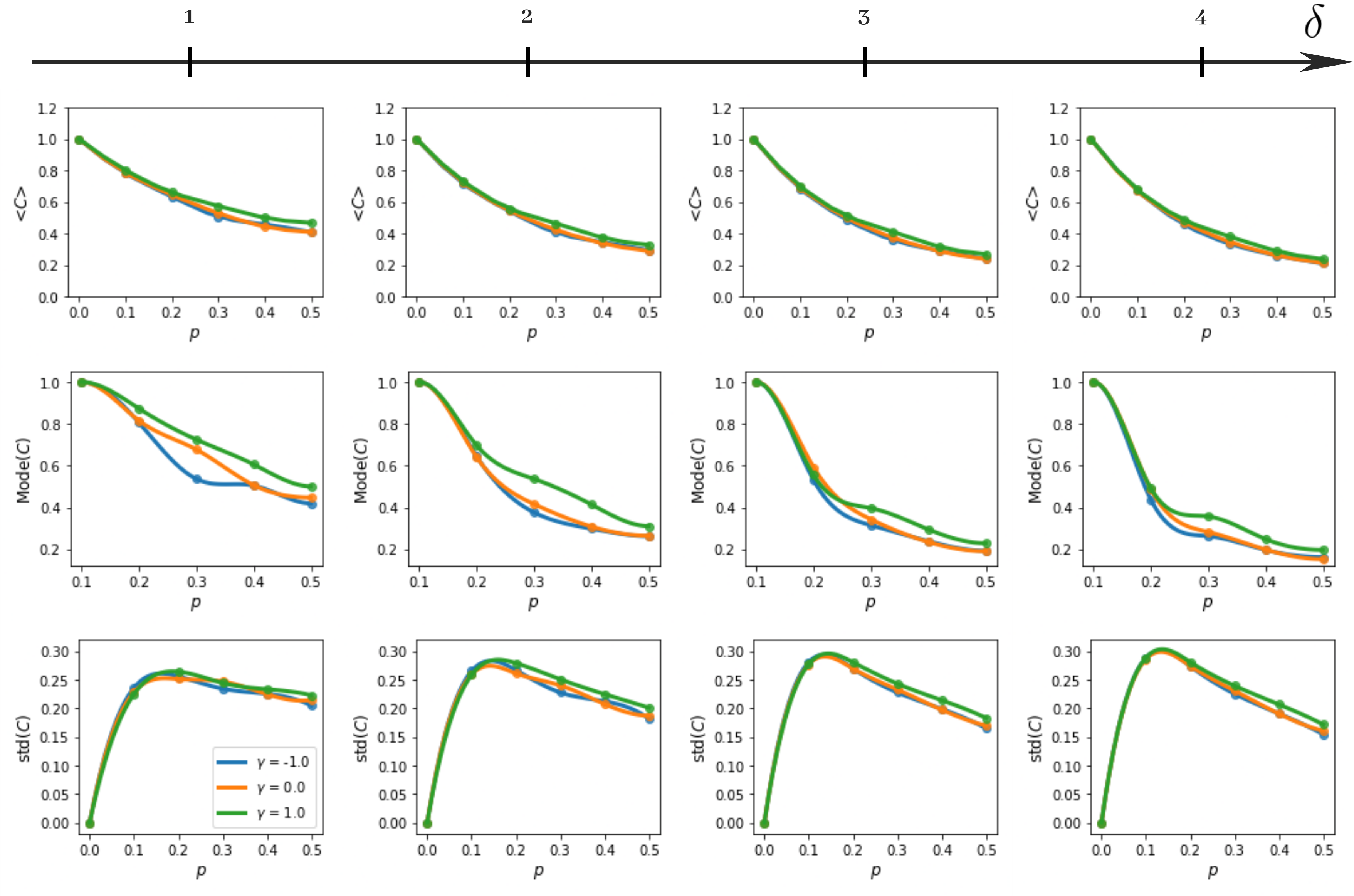}
  \caption{The average ($\left<C\right>$), mode ($Mode(C)$), and standard deviation ($std(C)$) of the coincidence values obtained for reconstructions of hierarchies with $N=15$ nodes and $\delta = 1, 2, 3, 4$, in terms of $\gamma=-1, -0, 1$ and $p=0.0, 0.1, 0.2, 0.3, 0.4, 0.5$.  As could be expected, all these measurements vary significantly with the probability error $p$. See text for respective discussion.}
  \label{fig:mean_std}
\end{figure*}

Figure~\ref{fig:mean_std} presents the average, mode, and standard deviation of the coincidence values obtained for reconstructions considering $N=15$ and $\delta=1, 2, 3, 4$ respectively to $\gamma = -1, 0,  1$ and $p=0.0, 0.1, 0.2, 0.3, 0.4, 0.5$. 

All three measurements can be verified to vary markedly with $p$. Relatively moderate differences can be observed between the average and standard deviation curves obtained respectively to the considered $\delta$ and$\gamma$ values, except for a larger decay ratio observed for larger values of $\delta$ verified in these two cases. Interestingly, the mode values present a stronger dependence on $\gamma$ and $\delta$. Generally speaking, it could be concluded that the mode of the obtained coincidence similarity indices vary with $\delta$, $\gamma$, and $p$, while the average and standard deviation are substantially less dependent on $\delta$ and $\gamma$.

\begin{figure*}[!ht]
  \centering
    \includegraphics[width=1\textwidth]{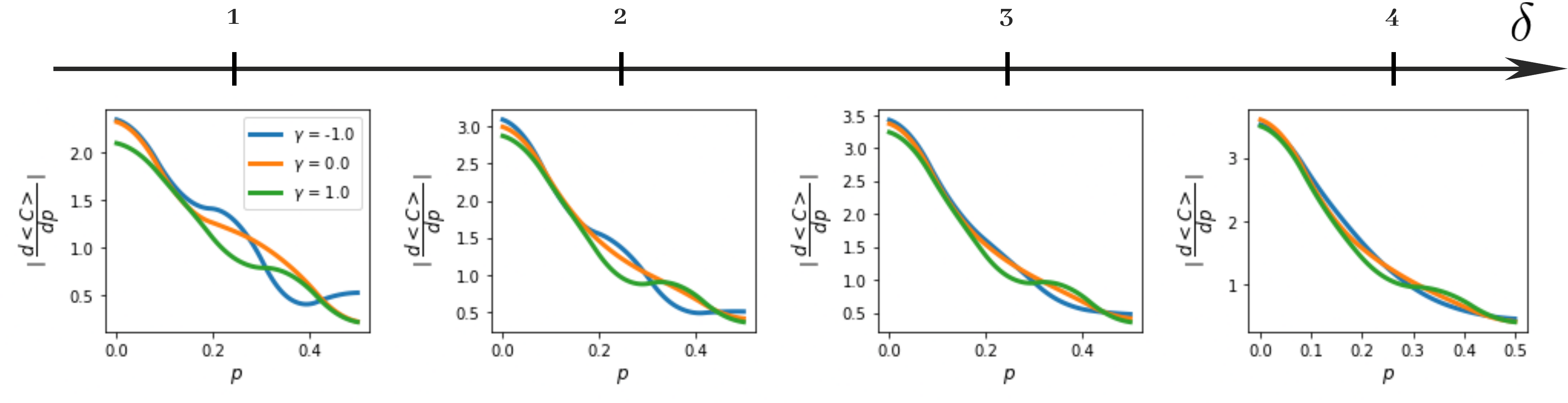}
  \caption{The sensitivity of the variation of the average reconstruction accuracy for $\gamma=-1, 0, 1$ and $\delta=1, 2, 3, 4$ in terms of the error probability $p$, as quantified by the absolute value of the derivative of the average coincidence values with respect to the error probability $p$ is substantially higher for small values of $p$, decreasing markedly as $p$ is increased.}
  \label{fig:Derivate}
\end{figure*}

Another interesting result concerns the fact that the largest decrease of average coincidence is observed, in all considered cases, along with the smallest values of $p$, tending to become substantially smaller for larger values of $p$. This type of effect, related to the sensitivity of the average of the obtained coincidence values, can be approached objectively in terms of the derivative of this measurement with respect to $p$. Figure~\ref{fig:Derivate} depicts this sensitivity with respect to $\gamma=-1,0,1$ and $\delta=1, 2, 3, 4$. This result tends to corroborate the above observation that relatively larger variations of the average reconstruction accuracy are obtained for the smallest values of $p$, decreasing as the latter parameter is increased.

\section{Concluding Remarks}\label{Concluding Remarks}

Several real-world structures and phenomena are characterized by respective hierarchical organization, to the point of being typically represented by respective trees. Examples of these situations include vascularization, neuronal cells, archaeological chronology, and phylogenetics, among many other possibilities. Even in structures not corresponding directly to trees, methods have been proposed to derive a respective hierarchical summarization, such as the minimal spanning tree (e.g.~\cite{dussert1986minimal}).

In practice, the acquisition of these structures often proceeds by sampling the tree nodes in a given order.  The sampled nodes are often appended to the currently available nodes while considering the similarity of their respective properties or features.  Given that the sampling order may not correspond to that originally characterizing the hierarchy, substantial errors can be verified in the respectively reconstructed structures.  

The present work focused on characterizing and studying the effect of the sampling order of hierarchical structures and phenomena respectively to several involved parameters, including the probability of error ($p$), the error extent ($\delta$), as well as the branching level of the respectively reconstructed structures ($\gamma$).  

In order to allow a systematic experimental investigation of the effect of these parameters on the obtained reconstruction errors, we developed a simple model for generating trees with varying branching levels, which are controlled by the parameter $\gamma$.  The trees constructed by this tree generating model were characterized by respective properties including the distributions of the number of hierarchical levels and the number of nodes per level. It has been verified that quite diverse types of trees can be obtained by the proposed model by varying its single parameter $\gamma$.  In addition to enabling the present study, this same model can be employed in several other applications.

The comparison between the original and reconstructed trees was quantified in terms of the coincidence similarity, which tends to provide a particularly strict quantification of the similarity between generic mathematical structures including the adjacency matrices used to represent the trees.

Several interesting results have been obtained and discussed.  These include the fact that, at least for the adopted parameter configurations and types of hierarchies, the reconstruction average accuracy varied relatively little with respect to both $\delta$ and $\gamma$, but decreased monotonically with $p$.  The mode of the coincidence values, however, presented a stronger variation with $\delta$ and $\gamma$.   In other words, retrieving hierarchies, as seen from the respective average and standard deviation of the obtained accuracies quantified by the respective coincidence values, is more critical on the probability error $p$ than on the type of hierarchy (indexed by $\gamma$) or the extent of the sampling order error $\delta$.
In addition, the relative variation of the accurac (sensitivity) was found to be substantially larger for smaller values of $p$, decreasing substantially for larger respective values.

The reported concepts, methods, and results paved the way for several related developments. To begin with, the proposed simple method for generating hierarchies with varying properties by using a single parameter can be adopted in several alternative problems and studies. Also worth investigating is the possibility of using alternative probability formulas dependent on $h$ and $k$ as the means for obtaining less overlap between trees respectively to distinct parameter configurations. Concerning the study of the reconstruction of hierarchies by sampling nodes, it would be interesting to study other types of sampling schemes and respective errors, as well as adopting other approaches for defining the respective features.

\section*{Acknowledgments}
Alexandre Benatti thanks Coordenação de Aperfeiçoamento de Pessoal de N\'ivel Superior - Brasil (CAPES) - Finance Code 001. Luciano da F. Costa thanks CNPq(grant no. 307085/2018-0) and FAPESP (grant 15/22308-2).

\bibliography{mybib}
\bibliographystyle{unsrt}

\end{document}